\documentclass[twoside,11pt]{article}

\usepackage{obs_study_style}
\usepackage{amsmath, bm}
\usepackage{graphicx}
\graphicspath{{figures/}} 
\usepackage{caption}
\usepackage{subcaption}
\usepackage[space]{grffile}
\usepackage{multirow}
\usepackage{booktabs}

\def\sumn{\sum_{i=1}^n} 
\def\prodn{\prod_{i=1}^n} 
\def\ipw{\textup{ipw}}
\def\reg{\textup{reg}}
\def\dr{\textup{dr}}
\def\hattauipw{\hat{\tau}^{\ipw}}
\def\hattaureg{\hat{\tau}^{\reg}}
\def\hattaudr{\hat{\tau}^{\dr}}
\def\ppp{\textup{PPP}}
\def\ind{\begin{picture}(9,8)
         \put(0,0){\line(1,0){9}}
         \put(3,0){\line(0,1){8}}
         \put(6,0){\line(0,1){8}}
         \end{picture}
        }

\def\diff{\textup{d}}
\newcommand{\N}{\mathcal{N}}

\firstpageno{1}

\begin{document}

\title{Posterior Predictive Propensity Scores and $p$-Values}


\author{\name Peng Ding \email pengdingpku@berkeley.edu \\
       \addr Department of Statistics\\
        University of California \\
        Berkeley CA 94720 U.S.A. \\
        \name Tianyu Guo \email guotianyu@pku.edu.cn \\
        \addr School of Mathematical Sciences\\
         Peking University \\
         No.5 Yiheyuan Road Beijing 100871 P.R.China}
\maketitle

\begin{abstract}
\citet{Rosenbaum83ps} introduced the notion of the propensity score and discussed its central role in causal inference with observational studies. Their paper, however, caused a fundamental incoherence with an early paper by  \citet{Rubin78}, which showed that the propensity score does not play any role in the Bayesian analysis of unconfounded observational studies if the priors on the propensity score and outcome models are independent. Despite the serious efforts made in the literature, it is generally difficult to reconcile these contradicting results. We offer a simple approach to incorporating the propensity score in Bayesian causal inference based on the posterior predictive $p$-value. To motivate a simple procedure, we focus on the model with the strong null hypothesis of no causal effects for any units whatsoever. Computationally, the proposed posterior predictive $p$-value equals the classic $p$-value based on the Fisher randomization test averaged over the posterior predictive distribution of the propensity score. Moreover, using the studentized doubly robust estimator as the test statistic, the proposed $p$-value inherits the doubly robust property and is also asymptotically valid for testing the weak null hypothesis of zero average causal effect. Perhaps surprisingly, this Bayesianly motivated $p$-value can have better frequentist's finite-sample performance than the frequentist's $p$-value based on the asymptotic approximation especially when the propensity scores can take extreme values. 
\end{abstract}

\begin{keywords}
Bayesian causal inference; doubly robust; frequentist's property; observational study; pivotal quantity; randomization test
\end{keywords}

\section{Causal inference with observational studies}
\label{sec::causal-pscore-central}

We focus on the canonical setting of causal inference with observational studies. We assume exchangeability of the units and thus drop the index $i$ for the $i$th unit $(i=1, \ldots, n)$. Let $Z$ denote the binary treatment, $Y$ denote the outcome of interest, and $X$ denote the pretreatment covariates. Under the potential outcomes framework \citep{Neyman23, Rubin74, Rosenbaum83ps}, let $Y(1)$ and $Y(0)$ denote the hypothetical outcome under the treatment and control, respectively. This framework allows us to define individual causal effect $Y(1) - Y(0)$ and the average causal effect 
$$
\tau =  E\{  Y(1) - Y(0) \} .
$$  
A fundamental difficulty of causal inference is that we cannot simultaneously observe both $Y(1)$ and $Y(0)$ for the same unit. The observed outcome $Y = ZY(1) + (1-Z)Y(0)$ equals $Y(1)$ for a treated unit with $Z=1$ and $Y(0)$ for a control unit with $Z=0$, respectively.

Following \citet{Rosenbaum83ps}, we assume unconfoundedness and overlap throughout: 
\begin{equation}
\label{condition::ignorability}
Z \ind \{  Y (1), Y (0) \} \mid X
\quad
\text{ and }
\quad 
0< e(X) <1 ,
\end{equation}
where 
$$
e(X) = P(Z=1\mid X)
$$ 
is the propensity score (PS). Under \eqref{condition::ignorability}, the average causal effect can be identified by
\begin{eqnarray}
\tau &=& E\left\{   \frac{ZY}{e(X)} - \frac{(1-Z)Y}{1-e(X)}  \right\}  \label{eq::ipw-identification} \\
&=& E\{ \mu_1(X) - \mu_0(X)  \} \label{eq::outreg-identification}
\end{eqnarray} 
where $ \mu_z(X)  = E( Y \mid  Z=z,X)$ is the outcome mean conditional on covariates under treatment $z$ $(z=0,1)$. 
The identification formula \eqref{eq::ipw-identification} motivates the inverse PS weighting estimator \citep{rosenbaum1987model}
$$
\hattauipw = \frac{\sumn Z_iY_i/\hat{e}(X_i)  }{ \sumn Z_i/\hat{e}(X_i) }
- \frac{\sumn (1-Z_i) Y_i/\{ 1-\hat{e}(X_i) \}  }{ \sumn (1-Z_i)/\{1-\hat{e}(X_i)\} }
$$ 
where $\hat{e}(X_i)$ denotes the estimated PS for unit $i$. Here we use the Hajek form instead of the Horvitz--Thompson form due to its superior finite-sample properties \citep{Lunceford04}. The identification formula \eqref{eq::outreg-identification} motivates the outcome regression estimator
$$
\hattaureg = n^{-1} \sumn \{ \hat \mu_1(X_i) -  \hat \mu_0(X_i) \}
$$
where $\hat \mu_1(X_i) $ and $ \hat \mu_0(X_i)$ denote the estimated conditional means of the outcomes under the treatment and control, respectively. The estimator $\hattauipw $ is consistent for $\tau$ if the PS model is correct, whereas the estimator $\hattaureg $ is consistent if the outcome model is correct. Motivated by the semiparametric efficiency theory \citep{bickel1998efficient, tsiatis2006semiparametric}, the doubly robust estimator combines both models \citep{Bang05}:
\begin{equation}
\label{eq::dr-estimator}
\hattaudr = \hattaureg  +  n^{-1} \sumn \left\{  \frac{Z_i R_i}{\hat{e}(X_i)} -  \frac{(1-Z_i) R_i}{1-\hat{e}(X_i)}   \right\}
\end{equation}
where the residual from outcome modeling $R_i = Y_i - \hat \mu_{Z_i}(X_i)$ equals $R_i = Y_i - \hat \mu_1(X_i)$ for a treated unit with $Z_i=1$ and $R_i = Y_i - \hat \mu_0(X_i)$ for a control unit with $Z_i=0$, respectively. The estimator $\hattaudr $ is consistent if either the PS or the outcome model is correct, justifying its name ``doubly robust.''

With parametric PS and outcome models, it is straightforward to construct estimators for the variances of these estimators based on the theory of M-estimation or the nonparametric bootstrap; see \citet{Lunceford04} and \citet{yang2020combining} for reviews. The recent literature has also extended these estimators to allow for more flexible nonparametric or machine learning estimation of the outcome model \citep{hahn1998role}, or the PS model \citep{Hirano03}, or both \citep{chernozhukov2018double}. We focus on estimators based on parametric models but conjecture that similar results extend to estimators based on more flexible models under some regularity conditions.

\section{The role of the propensity score in Bayesian causal inference}
\label{sec::propensityscore-ignorable-bayes}

\subsection{The propensity score is ignorable in Bayesian causal inference}
\label{sec::pscore-no-role}

Let $\theta_X$, $\theta_Z$, and $\theta_Y$ represent, respectively, the parameters for the models for the covariate distribution, the PS, the outcome conditional on the treatment and covariates. 
Rewrite the identification formula \eqref{eq::outreg-identification} of $\tau$ as
$$
\tau = \int  \{ \mu_1(x; \theta_Y) - \mu_0(x; \theta_Y) \}  f ( x; \theta_X) \text{d} x 
$$
which depends only on the unknown parameters  $\theta_X$ and $\theta_Y$. Assuming independent priors on the parameters $\theta_X$, $\theta_Z$ and $\theta_Y$, the joint posterior distribution based on exchangeable data $(X_i, Z_i, Y_i)_{i=1}^n$ factors into three independent components:
\begin{eqnarray*}
&&P( \theta_X, \theta_Z, \theta_Y\mid \cdot ) \\
&\propto & P(\theta_X)  \prodn P(X_i; \theta_X)   
 \cdot    P(\theta_Z) \prod_{i=1}^n  P(Z_i \mid X_i; \theta_Z)   
 \cdot   P(\theta_Y)  \prod_{i=1}^n  P(Y_i\mid Z_i, X_i; \theta_Y)
\end{eqnarray*}
The posterior distributions of $\theta_X$ and $\theta_Y$ do not depend on the second component corresponding to the PS. 
Therefore, Bayesian inference of $\tau$ does not depend on the PS. \citet{saarela2016bayesian} gave a similar discussion as above.

One might wonder whether the conclusions above will change if we use the identification formula \eqref{eq::ipw-identification} based on the inverse PS weighting. We can verify that under \eqref{condition::ignorability}, the formula \eqref{eq::ipw-identification} reduces to the formula \eqref{eq::outreg-identification}. The PS again does not play any role in Bayesian causal inference.

The above discussion focuses on $\tau$, the average causal effect of a super population. \citet{Rubin78} focused on the finite-sample average causal effect
$$
\tau_\text{fs} = n^{-1} \sum_{i=1}^n \{  Y_i(1)  - Y_i(0) \} ,
$$
and reduced the problem of causal inference to imputing the missing potential outcomes based on their posterior predictive distributions. Because $P(Y_i\mid Z_i, X_i; \theta_Y)$ depends only on $\theta_Y$, the PS can also be ignored in the finite-sample Bayesian causal inference.   By \citet{Rubin78}, the PS is {\it  ignorable}. \citet{Hill11} and \citet{Ding2018causalinference} discussed other parameters and reached the same conclusion.

\subsection{Existing strategies to use the propensity score in Bayesian causal inference}

The PS is central in frequentist's causal inference. Section \ref{sec::causal-pscore-central} above reviews its role in constructing the inverse PS and doubly robust estimators. In contrast, Section \ref{sec::propensityscore-ignorable-bayes} dismisses the role of the PS in Bayesian causal inference. A parallel discussions appeared in survey sampling \citep{Rubin85, pfeffermann1993role}.

Nevertheless, completely ignoring the PS seems worrisome. Because the PS characterizes the treatment assignment mechanism, it is intuitive to use it in one way or another in analyzing observational data. Below I will review some strategies to use the PS in Bayesian causal inference, with the last one being the proposal of this article. 

\paragraph{Use the PS in the design phase}
\citet{Rubin85} provided a heuristic argument based on robustness for the importance of using the PS in Bayesian causal inference. \citet{robins1997toward} provided more theoretical discussion of this issue. \citet{Rubin07} later argued that observational studies should have two stages: the design stage and the analysis stage. Based on this view, even the analysis stage is purely Bayesian in the sense of Section \ref{sec::pscore-no-role}, the PS plays a central role in the design stage to make the observational study as close as possible to a randomized experiment. Canonical approaches include PS matching and stratification in the design stage \citep{Rosenbaum83ps, Rubin07, ImbensRubin15}. This view highlights the role of the PS in designing observational studies but still cannot incorporate the PS in the Bayesian analysis reviewed in Section \ref{sec::propensityscore-ignorable-bayes}.

\paragraph{Use dependent priors}
Section \ref{sec::propensityscore-ignorable-bayes} assumes independent priors on the parameters $\theta_X$, $\theta_Z$ and $\theta_Y$. Consequently, the joint posterior distribution factors into three independent components, and then the PS is ignorable for inferring $\tau$. The independence of the posterior distributions, however, does not hold with dependent priors on $\theta_X$, $\theta_Z$ and $\theta_Y$. \citet{Wang12} used dependent prior for variable selection in both the PS and outcome models. \citet{ritov2014bayesian} constructed a dependent prior that yielded frequentist's properties. Their prior for the outcome model depended on the PS, and they only focused on some special cases. In general, this strategy may not be easy to implement to achieve desired frequentist's properties.

\paragraph{Use the PS as a covariate in the outcome model}
\citet{Zigler13}, \citet{an20104}, \citet{ZiglerDominici14}, \citet{zigler2016central}, and \citet{hahn2020bayesian} forced the PS to enter the outcome model in Bayesian computation. However, this strategy may be controversial. Arguably, the outcome model that reflects the natural of the potential outcomes generating process should not be dependent on the PS model that reflects the treatment assignment mechanism. Overall, while this strategy can be useful to improve robustness of causal inference, it relies on a somewhat unnatural factorization of the joint likelihood.

\paragraph{Posterior predictive estimation}
Based on the Bayesian posterior predictive perspective, \citet{saarela2016bayesian} proposed to use the posterior distribution of the doubly robust estimator $\hattaudr$ in \eqref{eq::dr-estimator}, with $\hat{e}(X_i)$ and $\{ \hat{\mu}_1(X_i), \hat{\mu}_0(X_i) \}$ drawn from their posterior distributions. \citet{antonelli2020causal} extended this idea to the setting with high dimensional covariates. This is a powerful idea to integrate frequentist's procedures in Bayesian causal inference.

\paragraph{Posterior predictive $p$-value}
Closely related to \citet{saarela2016bayesian} and \citet{antonelli2020causal} reviewed above, the proposal in the next section is based on the posterior predictive $p$-value (PPP). To motivate a simple procedure, we focus on the model of the strong null hypothesis of no causal effects for any units whatsoever. The resulting PPP is a natural extension of the classic Fisher randomization test (FRT) developed for randomized experiments. In observational studies, the proposed PPP equals the $p$-value based on the FRT averaged over the posterior predictive distribution of the PS. We present the details below.

\section{The PPP depends on the propensity score}
\label{sec::use-pscore-bayesian-ppp}

\subsection{General formulation of the PPP}
\label{sec::general-ppp}

We first show that the PS naturally enters Bayesian PPP. For simplicity, we focus on the model with the strong null hypothesis \citep{Rubin80}:
$$
H_{0\textsc{f}}: Y_i(1) = Y_i(0 ) = Y_i \text{ for all }i.
$$
The PS plays a central role in the PPP although it is ignorable in standard Bayesian inference reviewed in Section \ref{sec::pscore-no-role}. We will give a general formulation of the PPP for $H_{0\textsc{f}}$  below.

Focus on the finite samples at hand. Under the strong null hypothesis $H_{0\textsc{f}}$, the covariates and outcomes are all fixed, and the only random component is the treatment indicators. Under \eqref{condition::ignorability},  the posterior distribution of $\theta_Z$ is 
$$
P(\theta_Z\mid \cdot ) \propto 
P(\theta_Z )   \prod_{i=1}^n P(Z_i \mid X_i, Y_i; \theta_Z)     
= P(\theta_Z )    
\prod_{i=1}^n  P(Z_i \mid X_i; \theta_Z)  . 
$$
It only depends on the PS model  and reduces to a basic problem in Bayesian modeling. For instance, if $ P(Z_i \mid X_i; \theta_Z) $ follows a logistic model, then $P(\theta_Z\mid \cdot ) $ is the corresponding posterior distribution, which can be easily obtained using the \texttt{MCMClogit} function in the \texttt{MCMCpack} package in \texttt{R} \citep{martin2011mcmcpack}.

Define the statistic as $T= T(\bm Z, \bm X, \bm Y)$, where $\bm Z, \bm X, \bm Y$ are the concatenated treatments, covariates, outcomes for all observed units. 
Define 
$$
\ppp = P^\textup{pred}\left\{   T(\bm Z^\textup{pred}, \bm X, \bm Y)  \geq    T(\bm Z, \bm X, \bm Y) \right\}
$$
where $ P^\textup{pred}$ is the probability measure over the posterior predictive distribution of $\bm Z^\textup{pred}$ given the data:
$$
P^\textup{pred}(\bm Z^\textup{pred} \mid \cdot)
=   \int   \prod_{i=1}^n  P( Z_i^\textup{pred}\mid X_i ; \theta_Z) P( \theta_Z \mid \cdot) \diff \theta_Z ,
$$
where $ \bm Z^\textup{pred}  = (Z_1^\textup{pred}, \ldots, Z_n^\textup{pred}) $.  In this definition of the PPP, we consider a one-sided test. This PPP measures the tail probability of the test statistic $T(\bm Z, \bm X, \bm Y)$ compared to its posterior predictive distribution $T(\bm Z^\textup{pred}, \bm X, \bm Y)  $. By definition, the PS plays a central role in the Bayesian PPP.

The PPP was proposed for general Bayesian inference \citep{Rubin84, meng1994posterior, Gelman96}. 
It has also been applied to many  Bayesian causal inference problems \citep[e.g.,][]{Rubin84, rubin1998more, mattei2013exploiting, espinosa2016bayesian, jiang2016principal, forastiere2018posterior, zeng2020being}.

\subsection{Implementation of the PPP}\label{sec::implementation-ppp}

We then show how to implement the generic PPP introduced above. The first implementation follows the definition of the PPP closely: we simulate the test statistic by first drawing $\theta_Z$ from its posterior distribution and then drawing the treatment indicators conditional on $\theta_Z$. The detailed algorithm is below:
\begin{enumerate}
\item[A1]
draw $\theta_Z^r$ based on $P(\theta_Z \mid \cdot ) $,
draw $Z_i^r$ based on $P(Z_i = 1\mid X_i, \theta_Z^r )$ for all units $i=1,\ldots, n$, and
compute the statistic $T^r = T^r(\bm Z^r, \bm X, \bm Y)$;
\item[A2]
repeat the above step to obtain $T^r\ (r=1, \ldots, R)$;  
\item[A3]
calculate  
\begin{eqnarray}\label{eq::ppp-mc}
\ppp \stackrel{\cdot}{=}  R^{-1} \sum_{r=1}^R \mathcal{I}( T^r  \geq  T  ).
\end{eqnarray} 
\end{enumerate}
Computationally, the above algorithm is straightforward. 
We use it to compute the PPP in simulation in Section \ref{sec::simulation-studies}.

Moreover, an alternative implementation below can provide more insights into the PPP. By swapping the integrals in the definition of the PPP, we can rewrite the PPP as the FRT averaged over the posterior predictive distribution of the PS. Mathematically,
\begin{eqnarray}\label{eq::ppp-def}
\ppp = \int p(\theta_Z )   P(\theta_Z \mid \cdot ) \diff \theta_Z
\end{eqnarray} 
where $p(\theta_Z) $ is the $p$-value for a fixed $\theta_Z$:
$$
p(\theta_Z )  = P\left\{   T(\bm Z^\textup{pred}, \bm X, \bm Y)  \geq    T(\bm Z, \bm X, \bm Y) \mid \theta_Z \right\}
$$
with the $  Z^\textup{pred}_i$'s drawn independently from $P ( Z_i^\textup{pred}\mid X_i ;  \theta_Z) $ $(i=1,\ldots, n)$. Given $\theta_Z$, we know the PS for all units and thus we effectively have a randomized experiment with varying treatment probabilities $P(Z_i=1\mid X_i; \theta_Z)$ across units. Therefore, $p(\theta_Z ) $ is the $p$-value from the standard FRT. We give the details for calculating $p(\theta_Z ) $ below: 
\begin{enumerate}
\item[B1]
draw $Z_i^s(\theta_Z)$ based on $P(Z_i = 1\mid X_i; \theta_Z)$ for all units $i=1,\ldots, n$, and
compute the statistic $ T^s(\theta_Z) = T^s(\bm Z^s(\theta_Z), \bm X, \bm Y)$;
\item [B2] repeat the above step to obtain $T^s(\theta_Z)$ $(s=1, \ldots, S)$;
\item[B3] 
calculate  
$$
p(\theta_Z) \stackrel{\cdot}{=} S^{-1} \sum_{s=1}^S \mathcal{I}( T^s(\theta_Z) \geq   T  ).
$$
\end{enumerate}
The PPP equals the average of $p(\theta_Z)$ over the posterior distribution of $\theta_Z$ by \eqref{eq::ppp-def}. \citet{meng1994posterior} and \citet{Gelman96} re-formulated the PPP as \eqref{eq::ppp-def} and motivated the above procedure B1--B3.

As a side comment, the FRT interpretation of the PPP is natural in causal inference. This interpretation cannot be generalized to the survey sampling setting although the existing literature focused more on the commonality of causal inference and survey sampling \citep[e.g.,][]{Rubin85, Bang05}. They differ in this aspect.

\subsection{Choice of the test statistic: studentized estimators are superior}

We now discuss the choice of the test statistic. The generic PPP introduced in Section \ref{sec::general-ppp} allows for using any test statistic. However, practitioners may find the strong null hypothesis restrictive. Moreover, frequentist's statisticians may completely dismiss the PPP due to its Bayesian nature. To address these concerns, we propose to use the studentized doubly robust statistic in the PPP, which yields an asymptotically valid $p$-value for the weak null hypothesis  
$$
H_{0\textsc{n}}:  \tau  = 0
$$
for the average causal effect. This guarantee is under the frequentist's paradigm \citep[cf.][]{robins2000asymptotic} even though the original PPP is motivated by Bayesian thinking.

Even though Section \ref{sec::causal-pscore-central} has a completely different motivation from the  Bayesian PPP, the estimators there can provide important insights into choosing the test statistic. If $H_{0\textsc{n}}$ is of interest, then intuitively,  we can choose $T$ as the absolute value of $\hattauipw$, $\hattaureg$ or $\hattaudr$. Previous results for the FRT \citep{chung2013exact, wu2021randomization, zhao2021covariate}, however, suggest that using them in the PPP does not ensure correct type one error rate even in large samples. Better choices are the absolute value of the {\it studentized} estimators, that is, $\hattauipw$, $\hattaureg$ and $\hattaudr$ divided by the corresponding consistent standard errors.

Our simulation below will demonstrate the superiority of the studentized estimators, especially the one based on the doubly robust estimator. We focus on the empirical evaluation of the PPP and leave the rigorous frequentist's theory to another report.

\subsection{Special case: the classic FRT}

With a randomized experiment, $P(Z_i, \ldots, Z_n \mid X_1,\ldots, X_n)$ is determined by the designer of the experiment without any unknown parameter. In this case, $\theta_Z$ is empty and we do not need to simulate its posterior distribution. In calculating the PPP, we simply simulate the treatment indicators from $P(Z_i, \ldots, Z_n \mid X_1,\ldots, X_n)$, following the same rule as the initial randomized experiment. This is precisely the FRT, as pointed out by \citet[][Section 5.6]{Rubin84} and \citet[][Section 4]{Rubin05}.

\citet{Rubin84, Rubin05} pointed out the Bayesian interpretation of the FRT and thus hinted at the idea in this article. However, he did not pursue the general form of the PPP proposed in Section \ref{sec::general-ppp} for observational studies.

\section{Simulation: frequentist's evaluation of the Bayesian PPP}
\label{sec::simulation-studies}

In this section, we evaluate the finite-sample performance of the PPP via simulation. The results suggest that the PPP using the studentized doubly robust estimator has the most desirable properties. 

For the frequentist's evaluation, we repeatedly generate the data for 3000 times. In each replication, we follow the procedures A1--A3 in Section \ref{sec::implementation-ppp} to calculate the PPP. We use the function \texttt{MCMClogit} in the \texttt{MCMCpack} to simulate the posterior distribution of the coefficients of the logistic PS model, with an improper uniform prior on $\theta_Z$, 1000 burn-in iterations, and 2000 draws of the $\theta_Z^r$'s. 

\subsection{Data generating process and model specification}

We choose the sample size   as $n=1000$. We consider two different types of data generating process (DGP). 

\paragraph{DGP without extreme PS}
We first generate four covariates from
\begin{eqnarray*}
&X_{i1}=W_{i1},\quad 
X_{i2}=W_{i2}+0.3X_{i1},&\\
&X_{i3}=W_{i3}+0.2(X_{i1}X_{i2} - X_{i2}),\quad 
X_{i4}=W_{i4} +0.1(X_{i1}+X_{i3}+X_{i2}X_{i3}),&
\end{eqnarray*}
with
$$
W_{i1} \sim \text{Bernoulli}(0.5),\quad 
W_{i2} \sim \text{Uniform}(0,2),\quad 
W_{i3} \sim \text{Exponential}(1),\quad 
W_{i4} \sim \chi^2(4).
$$ 
The PS follows the logistic model
$$
P(Z_i=1\mid X_i; \theta_Z) = \{ 1+\exp(-X_i^{\scriptscriptstyle\text  T}\theta_Z) \}^{-1} \quad 
\text{ with } \theta_Z = (-1,0.5,-0.25,-0.1)^{\scriptscriptstyle \text T} .
$$
The outcomes follow the linear models:
$$
    Y_i(0)= \ \mu_0+({X}_i - \mu)^{\scriptscriptstyle \text T} {\beta}_{1}+ \epsilon_i(0) , \quad
    Y_i(1)= \ \mu_1+ ({X}_i - \mu)^{\scriptscriptstyle \text T}{\beta}_{0}+ \epsilon_i(1), 
    $$
    with $\epsilon_i(0)\sim \N(0,5^2)$, $\epsilon_i(1) \sim \N(0,1^2)$, and 
$$
\mu_1=\mu_0=1, \quad \mu =  E  (X) ,\quad \beta_1 = (0.1,-0.2,-0.2,-0.2)^{\scriptscriptstyle \text T} ,\quad
    \beta_0 = (-0.1,0.3,0.1,-0.2)^{\scriptscriptstyle \text T} .
$$
So $\tau = E \{ Y(1) \} -  E \{  Y(0) \} = \mu_1 - \mu_0=0$. 

%
%
%

\paragraph{DGP with extreme PS}
We first generate two covariates
$$
X_{i1} = \exp(W_{i1}), \quad X_{i2} = \exp(W_{i2}) \quad \text{ with } (W_{i1},W_{i2})^{\scriptscriptstyle \text T} \sim \N(0,I_2). 
$$
The PS model is 
$$
P(Z_i=1\mid X_i; \theta_Z ) = \{ 1+\exp(1-X_i^{\scriptscriptstyle\text  T}\theta_Z) \}^{-1} 
\quad \text{ with } \theta_Z = (1,-1)^{\scriptscriptstyle \text T}.
$$
Change the coefficients of the outcome models  to 
$$
\mu_0 = \mu_1 = -1 + 0.1\sqrt{e}, \quad 
\beta_1=(-0.2, 0.1)^{\scriptscriptstyle\text  T}, \quad 
\beta_0=(0.2,-0.1)^{\scriptscriptstyle\text  T}.  $$ 
So again $\tau = 0$.

For each DGP, we consider four combinations of model specifications:

\begin{enumerate}
\item[(i)] 
Both the PS and the outcome models are correctly specified.

\item[(ii)] 
The PS model is correctly specified but the outcome model is misspecified. In particular, for the DGP without extreme PS, we regress $Y$ on $W_2$ and $W_3$; for the DGP with extreme PS, we regress $Y$ on $W_1$ and $W_2$.

\item[(iii)] 
The outcome model is correctly specified but the PS model is misspecified. In particular, for the DGP without extreme PS, we regress $Z$ on $W_2$ and $W_3$; for the DGP with extreme PS, we regress $Z$ on $W_1$ and $W_2$.

\item[(iv)] 
Both the PS and the outcome models are misspecified. 
\end{enumerate}


In this article, we focus on simulation with low dimensional covariates and parametric models. Extending to simulation with high dimensional covariates and more flexible models requires more sophisticated choices of the priors. This is an important future research direction.

\subsection{Simulation under the weak null hypothesis}

\begin{figure}[t]
\centering
\includegraphics[width = 0.85\textwidth]{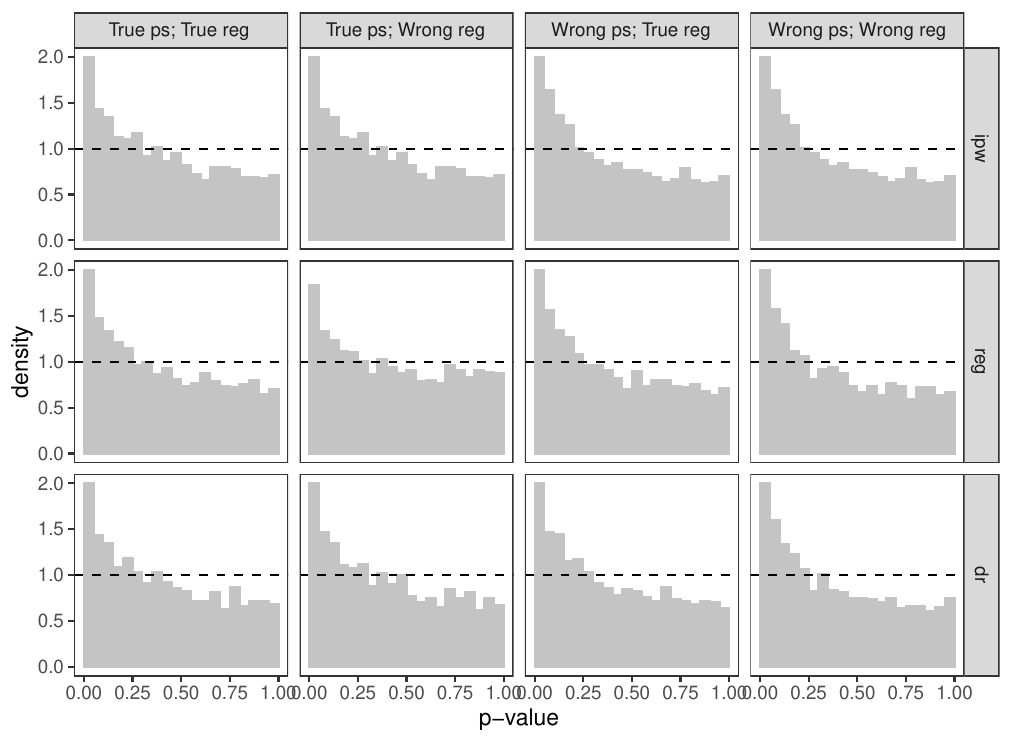}
\caption{PPP using the unstudentized test statistics under $H_{0\textsc{n}}$ and the DGP without extreme PS.
To obtain the anti-conservative PPP, we change $Z_i$ to $1-Z_i$. The densities are truncated at $2$.}

\label{weak null unstu}
\end{figure}

We first show that the problem of using the unstudentized statistics under the weak null hypothesis $H_{0\textsc{n}}$.  The original DGP without extreme PS yields conservative PPP. Once we flip the treatment and control group indicators, we can get anti-conservative PPP. See Figure \ref{weak null unstu}.  

\begin{figure}[t]
\centering
\includegraphics[width = 0.85\textwidth]{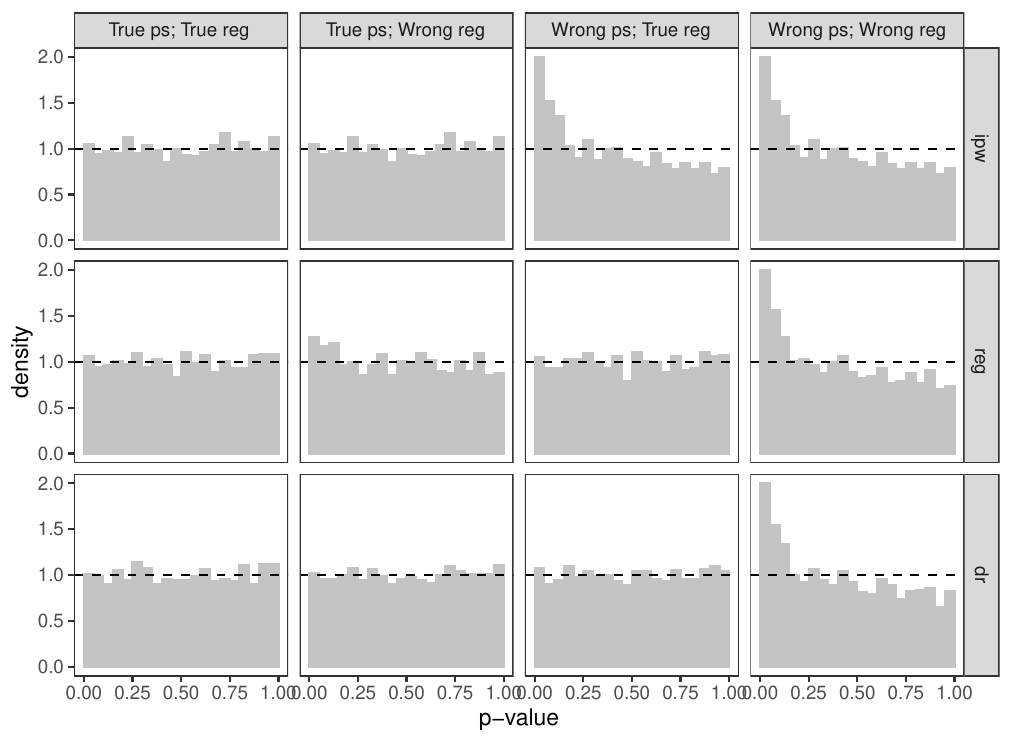}

(a) DGP without extreme PS 

\includegraphics[width = 0.85\textwidth]{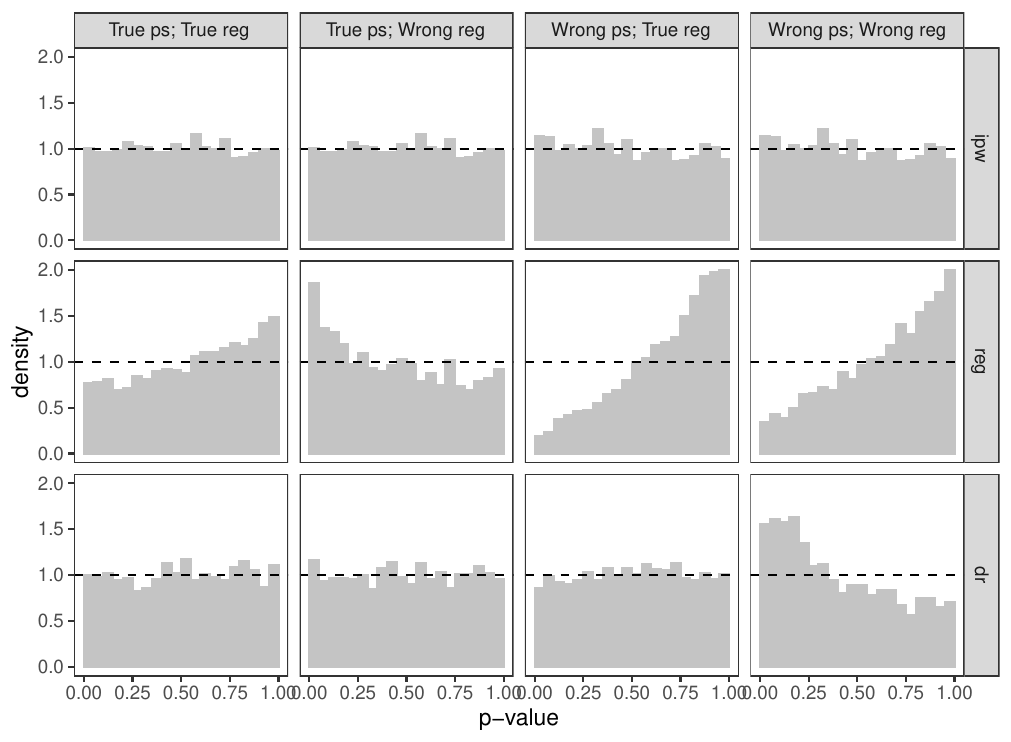}

(b) DGP with extreme PS

\caption{PPP using studentized test statistics under $H_{0\textsc{n}}$. The densities are truncated at $2$. }\label{fig::studentization}
\end{figure}

We then show the superiority of using the studentized statistics under weak null hypothesis $H_{0\textsc{n}}$. For computational simplicity, we use the estimated standard errors based on the theory of M-estimation. Figure \ref{fig::studentization}(a) shows the distribution of the PPP under the DGP without extreme PS. The PPP has uniform distributions with correctly specified models. The PPP with the studentized doubly robust estimator is doubly robust since it is uniform if either the PS or outcome model is correctly specified. It is our final recommendation.

Under the DGP with extreme PS, the superiority of our recommendation becomes clearer. See Figure \ref{fig::studentization}(b).

\subsection{Comparison of the PPP with normal approximation}

We now compare the performance of PPP and the normal approximation based on the studentized doubly robust estimator. We use the standard errors based on both the asymptotic expansion and the bootstrap by resampling the data 2000 times. So in total, we compare four $p$-values.

\begin{figure}[t]
\centering
\includegraphics[width = \textwidth]{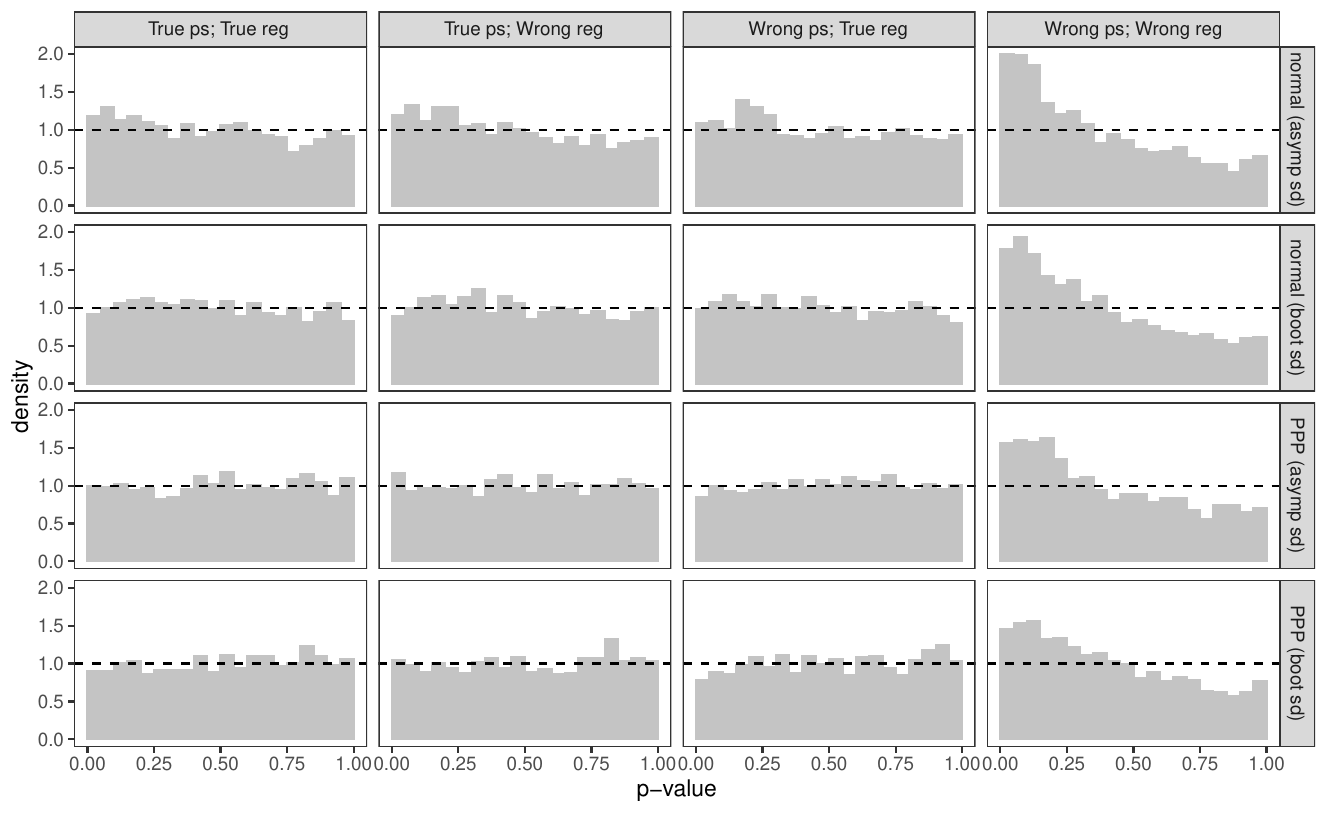}

(a) under the DGP with extreme PS and the weak null hypothesis

\includegraphics[width = \textwidth]{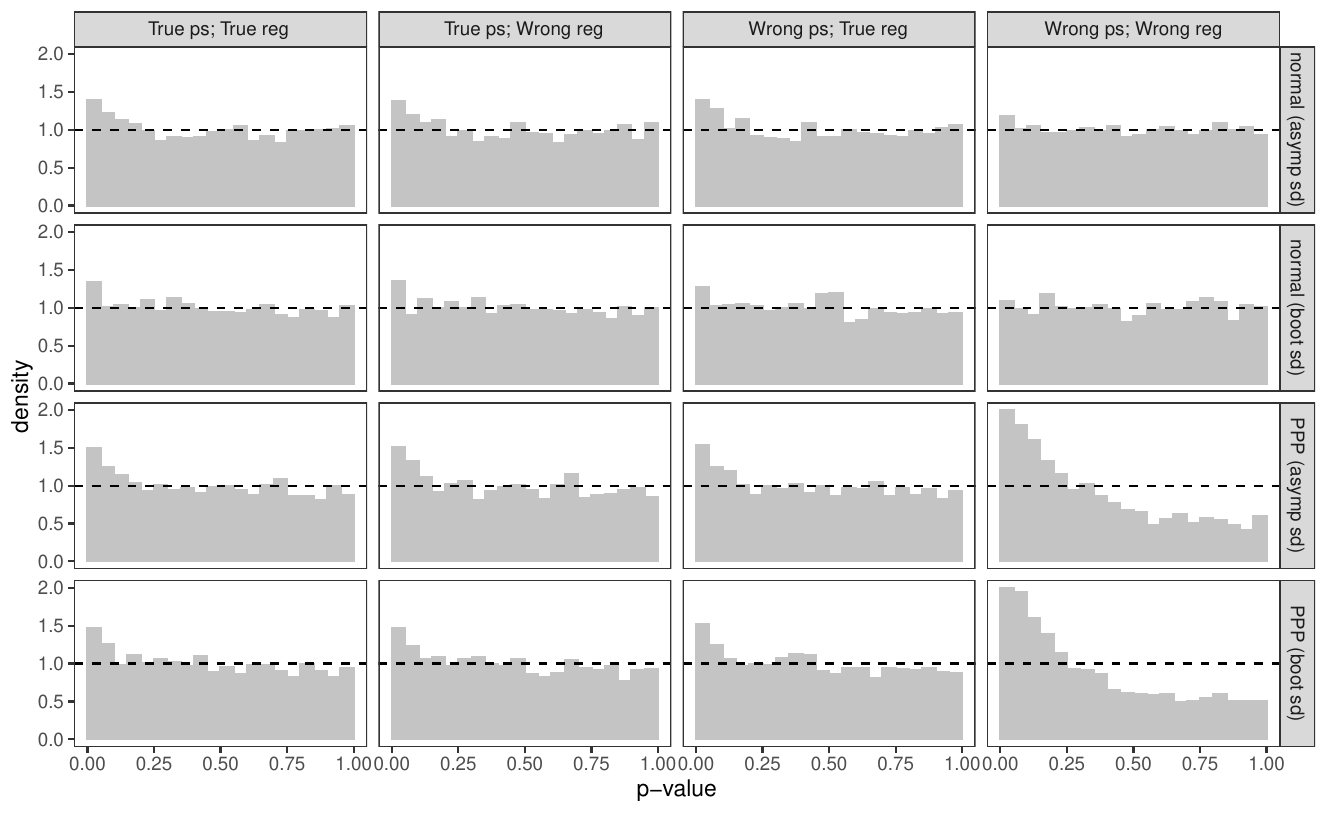}

(b) under the DGP without extreme PS and an alternative hypothesis

\caption{Comparison of the PPP with normal approximation based on the studentized doubly robust estimator. The densities are truncated at 2.}
\label{fig::ppp-normal-compare}
\end{figure}

We first compare them under the weak null hypothesis $H_{0\textsc{n}}$. 
Under the DGP without extreme PS, they have similar performance, so we omit the results. Under the DGP with extreme PS, the bootstrap or PPP alone has superior performance compared to the normal approximation based on the asymptotic standard error; their combination does not yield further improvement. Figure \ref{fig::ppp-normal-compare}(a) shows the histograms of the four $p$-values under four scenarios.

We then compare their power under an alternative hypothesis with a nonzero $\tau$. We use the DGP without extreme PS for this case. Let $\mu_1 = 1.1$ and $\mu_0 = 1$ so that $\tau = \mu_1-\mu_0 = 0.1$.  In this scenario, all $p$-values have similar power. See Figure \ref{fig::ppp-normal-compare}(b).

\subsection{Replication files and data analysis}

The replication files of this article can be found at Harvard Dataverse: 
\begin{quotation}
https://doi.org/10.7910/DVN/QPOS31
\end{quotation}
There we post the \texttt{R} code for the simulation studies as well as two data analysis examples.

\section{Discussion}

\subsection{Summary}

We first reviewed the conceptual difficulty of using the PS in Bayesian causal inference in Section  \ref{sec::propensityscore-ignorable-bayes}.  We then build upon \citet{Rubin84} to proposed a PPP in Section \ref{sec::use-pscore-bayesian-ppp}, which naturally uses the PS and extends the classic FRT by averaging over the posterior predictive distribution of the PS. Moreover, we recommend using the studentized doubly robust estimator in the PPP, which yields superior finite-sample properties even from the frequentist's perspective under the weak null hypothesis.

\subsection{Frequentist's properties}

We have used simulation in Section \ref{sec::simulation-studies} to evaluate the frequentist's properties of the PPP which leads to the following conjecture:

\noindent {\bfseries  Conjecture}: Assume $\tau = 0$ and regularity conditions. The PPP with the studentized doubly robust estimator, $\text{PPP}^\text{dr}$, has the following asymptotic property:
$$
\text{PPP}^\text{dr}  \stackrel{\text{d}}{\longrightarrow} \text{Uniform}(0,1) ,\qquad \text{ as }  n\rightarrow \infty
$$
if either the PS or the outcome model is correctly specified. 

The conjecture is a frequentist's statement although $\text{PPP}^\text{dr}$ itself is motivated by a Bayesian procedure. Intuitively, it holds because the studentized doubly robust estimator is asymptotically pivotal if either the PS or the outcome model is correctly specified. It ensures that we can use $\text{PPP}^\text{dr}$ as a standard frequentist's $p$-value for testing the weak null hypothesis of zero average treatment effect.  We leave the proof of the conjecture to future work.

\subsection{Epilogue: Did \citet{Rosenbaum83ps} mention the Bayesian PS?}

Yes, they did. In \citet[][Section 1.3]{Rosenbaum83ps}, they wrote:
\begin{quote}
To a Bayesian, estimates of these probabilities are posterior predictive probabilities of assignment to treatment $1$ for a unit with vector $x$ of covariates.
\end{quote}
However, they did not provide any further discussion on the role of the PS in Bayesian causal inference perhaps due to the incoherence with \citet{Rubin78}. The existing literature has clearly documented the difficulty of using the PS in standard Bayesian causal inference. We argue that the PPP is a natural approach to incorporate the PS in Bayesian causal inference. Under the model of the strong null hypothesis, it can be viewed as the FRT averaged over the posterior predictive distribution of the PS.

\acks{Peng Ding thanks the National Science Foundation (\# 1945136) for support, Fan Li for insightful discussion, Zhichao Jiang and Avi Feller for helpful comments. 
Tianyu Guo is partially supported by the elite undergraduate training program of School of Mathematical Sciences in Peking University.}





%
%

\vskip 0.2in
\bibliography{ppppp}

\end{document}